\documentclass[pra,showpacs,tighten,aps,amssymb,twocolumn,superscriptaddress]{revtex4}

\usepackage{times}
\usepackage{bbm}
\usepackage{epsf}

\newcommand{\naturals}{\mathbbm{N}}
\newcommand{\reals}{\mathbbm{R}}
\newcommand{\proofend}{\hfill\rule[0pt]{.2cm}{.2cm}\medskip }

\begin{document}
\title{Remarks on entanglement measures and non-local state distinguishability}
\author{J. Eisert}
\affiliation{QOLS, Blackett Laboratory, Imperial College London,
London SW7 2BW, UK}
\affiliation{Institut f{\"u}r Physik, University of Potsdam, D-14469
Potsdam, Germany}
\author{K. Audenaert}
\affiliation{QOLS, Blackett Laboratory, Imperial College London,
London SW7 2BW, UK}
\affiliation{School of Informatics, University
of Wales, Bangor LL57 1UT, UK}
\author{M.B. Plenio}
\affiliation{QOLS, Blackett Laboratory, Imperial College London,
London SW7 2BW, UK}
\date{\today}

\begin{abstract}
We investigate the properties of three entanglement measures that
quantify the statistical distinguishability of a given state with
the closest disentangled state that has the same reductions as the
primary state. In particular, we concentrate on the relative
entropy of entanglement with reversed entries. We show that this
quantity is an entanglement monotone which is strongly additive,
thereby demonstrating that monotonicity under local quantum
operations and strong additivity are compatible in principle. In
accordance with the presented statistical interpretation which is
provided, this entanglement monotone, however, has the property
that it diverges on pure states, with the consequence that it
cannot distinguish the degree of entanglement of different pure
states. We also prove that the relative entropy of entanglement
with respect to the set of disentangled states that have identical
reductions to the primary state is an entanglement monotone. We
finally investigate the trace-norm measure and demonstrate that it
is also a proper entanglement monotone.
\end{abstract}
\maketitle
\section{Introduction}
Quantum entanglement arises as a joint consequence of the
superposition principle and the tensor product structure of the
quantum mechanical state space of composite quantum
systems. One of the main concerns of a theory of quantum entanglement
is to find mathematical tools that are capable of
appropriately quantifying the extent to which composite
quantum systems are entangled. Entanglement measures
are functionals that are constructed to serve that
purpose
\cite{Horo1,Intro1,Bennett,Uniqueness,Rains,Wootters,Cost,Plenio,Relent2,Asy,Asy2,Mu,Simon,Neg,Limits,Vidal}.
Initially it was hoped for that a number of natural requirements
reflecting the properties of quantum entanglement would be
sufficient to establish a unique functional that quantifies
entanglement in bi-partite quantum systems \cite{Uniqueness}.
These requirements are the non-increase (monotonicity) of the
functional under local operations and classical communication, the
convexity of the functional (which amounts to stating that the
loss of classical information does not increase entanglement) and
the asymptotic continuity. Indeed, for pure quantum states these
contraints essentially define a unique measure of entanglement.
This uniqueness originates from the fact that pure-state
entanglement can asymptotically be manipulated in a reversible
manner \cite{Bennett} under local operations with classical
communication (LOCC). However, for mixed states there is no such
unique measure of entanglement, at least not under LOCC (see
however, \cite{PPT,Asy3}). Instead, it depends very much on the
physical task underlying the quantification procedure what degree
of entanglement is associated with a given state. The distillable
entanglement grasps the resource character of entanglement in
mathematical form: it states how many maximally entangled
two-qubit pairs can asymptotically be extracted from a supply of
identically prepared quantum systems \cite{Bennett,Rains}. The
entanglement of formation \cite{Bennett,Wootters}---or rather its
asymptotic version, the entanglement cost under LOCC
\cite{Cost,Cost2}---quantifies the number of maximally entangled
two-qubit pairs that are needed in an asymptotic preparation
procedure of a given state.

The relative entropy of entanglement
\cite{Plenio,Relent2,Asy,Asy2,Mu,Simon} is an intermediate
measure: it has an interpretation in terms of statistical
distinguishability of a given state of the closest 'disentangled'
state. This set of 'disentangled' states could be the set of
separable states, or the set of states with a positive partial
transpose (PPT states). The relative entropy of entanglement
quantifies, roughly speaking, to what minimal degree a machine
performing quantum measurements could tell the difference between
a given state and any disentangled state \cite{Plenio}.

It is not unthinkable that the optimal disentangled state may
already be distinguishable from the primary state using selective
local operations, rather than global ones. Yet, it would be
interesting to see what measures of entanglement would arise if
one considered only those disentangled states that can not be
distinguished locally from the primary state, specifically that
both states have identical reductions with respect to both parts
of the bi-partite quantum system. In this sense one asks for the
degree to which the two states can be distinguished in a genuinely
non-local manner.

It is the purpose of this paper to pursue this program. We will
discuss three different entanglement measures that are related to
this distinguishability problem. Each of these entanglement
measures is based on a different state space distance measure,
namely on the relative entropy, the relative entropy with
interchanged arguments and the trace-norm distance. The properties
of these entanglement measures have not been studied so far. We
will show that these three quantities are entanglement monotones,
thereby qualifying them as proper measures of entanglement.

An interesting byproduct of this work is the result that
the relative entropy of entanglement with interchanged arguments
is strongly additive, which means that
\begin{equation}
E(\sigma\otimes\rho)=E(\sigma)+E(\rho)
\end{equation}
for all states $\rho$ and $\sigma$. Strong additivity
implies weak additivity, i.e.\ $E(\rho^{\otimes n})= n E(\rho)$
for all states $\rho $ and all $n\in\naturals$.
If one can interpret an entanglement measure as a kind of cost function,
weak additivity can be interpreted as the impossibility to
get a 'wholesale discount' on a state.
Many measures of entanglement are known to be subadditive, such as the relative entropy
of entanglement and the non-asymptotic entanglement of formation.
Furthermore, all regularized asymptotic versions of entanglement measures
are, by definition, weakly additive.
As no strongly additive measure of entanglement
has been found so far, one might be led to doubt whether
the requirements of (i) monotonicity, (ii) strong additivity,
and (iii) convexity are compatible at all. We will show, however, that the relative entropy
of entanglement with interchanged arguments, and taken with respect to the set of disentangled
states with the same reductions as the primary state, obeys each one of these three requirements,
proving that there is no a priori incompatibility between them.
It has to be noted, though, that this result is of a rather technical nature, as this measure of
entanglement, while being physically meaningful, is not very practical: it yields infinity for any
pure entangled state.

\section{Notation and definitions}

In this work we will consider bi-partite systems consisting of
parts $A$ and $B$, each of which is equipped with a
finite-dimensional Hilbert space. The set of density operators of
the joint system will be denoted as ${\cal S}({\cal
H})$.
Let ${\cal D}({\cal H})$ be either the set of separable
states or the set of PPT states, which is the subset
of ${\cal S}({\cal H})$ which consists of the states $\sigma$
for which the partial transpose $\sigma^\Gamma$ is a positive operator.
In the following, we will consider the proper subset ${\cal D}_\sigma({\cal H})
\subset {\cal D} ({\cal H})$ which consists of all those separable states (or
PPT states) that are locally identical to $\sigma$,
\begin{equation}
    {\cal D}_\sigma({\cal H}):=
    \left\{
    \rho \in {\cal D}({\cal H}):
    \rho_A=\sigma_A, \rho_B=\sigma_B
    \right\}.
\end{equation}
In this definition, subscripts $A$ and $B$ denote state reductions to
the subsystems $A$ and $B$, respectively.
The quantities that will be considered in this paper
are all distance measures with respect to this set:
\begin{eqnarray}
    E_A(\sigma) &:=& \inf_{\rho\in {\cal D}_\sigma({\cal H})}
    S(\rho\|\sigma),\label{EA}\\
    E_M(\sigma) &:=& \inf_{\rho\in {\cal D}_\sigma({\cal H})}
    S(\sigma\|\rho),\label{ER}\\
    E_T(\sigma) &:=& \inf_{\rho\in {\cal D}_\sigma({\cal H})}
    \| \rho- \sigma\|_1,\label{ET}
\end{eqnarray}
where
\begin{equation}
S(\rho\|\sigma)=\text{tr}[\rho\log_2 \rho - \rho\log_2 \sigma]
\end{equation}
is the relative entropy \cite{Ohya,Wehrl}, and $\|.\|_1$ stands for the
trace norm \cite{Bhatia}.

The quantity $E_M$ in
Eq.\ (\ref{ER}) is the relative entropy of entanglement
\cite{Plenio,Relent2} of a state
$\sigma$ with respect to the set ${\cal D}_\sigma({\cal H})$. The
original relative entropy of entanglement with respect to the set
${\cal D}({\cal H})$ (meaning either separable or PPT states) is
an entanglement measure that has been extensively studied in the
literature \cite{Plenio,Relent2}. Initially formulated as a
quantity for bi-partite finite dimensional systems, it has later
been generalized to the asymptotic \cite{Asy}, the multi-partite
\cite{Mu}, and the infinite-dimensional setting \cite{Simon}.
$E_A$ in Eq.\ (\ref{EA})
is essentially
the relative entropy with reversed entries,
first mentioned in Ref.\ \cite{Plenio}. The particular property of
this quantity is that it is strongly additive. The quantity $E_T$
in Eq.\ (\ref{ET})
is a distance measure based on the trace norm.
All quantities are related to the minimal degree to
which a given bi-partite state $\sigma$ can be
distinguished from any state taken from ${\cal D}({\cal H})$
that cannot be distinguished by purely local means with
operations in $A$ or $B$ only. This statement will be made more
precise in Section VI.

The properties of $E_A$, $E_M$ and $E_T$ that will be
investigated consist of the following well-known list of (non-asymptotic)
properties of proper entanglement
measures  \cite{Bennett,Plenio,Limits,Vidal,Uniqueness}:

\begin{itemize}
\item[(i)] If $\sigma \in {\cal S}({\cal H})$
is separable, then $E(\sigma)=0$.
\item[(ii)]There exists a $\sigma\in {\cal S}({\cal H})$
for which $E(\sigma)>0$.
\item[(iii)] {\it Convexity}\/:
Mixing of states does not increase
entanglement: for all $\lambda\in[0,1]$ and all $\sigma_1,\sigma_2\in {\cal S}({\cal H})$
\begin{eqnarray}
    E(\lambda \sigma_1 + (1-\lambda)\sigma_2)\leq
    \lambda E(\sigma_1) + (1-\lambda) E(\sigma_2).
\end{eqnarray}
\item[(iv)] {\it Monotonicity under local operations}\/:
Entanglement cannot increase on average under
local operations: If one performs
a local operation in system $A$ leading to
states $\sigma_{i}$ with respective
probability $p_{i}$, $i=1,\ldots,N$, then
\begin{equation}\label{mon}
    E(\sigma)\geq  \sum_{i=1}^{N} p_i E(\sigma_{i}).
\end{equation}
\item[(v)] {\it Strong
additivity}\/: Let ${\cal H}$ have the structure
${\cal H}^{(1)}\otimes {\cal H}^{(2)}$, with
\begin{eqnarray}
    {\cal H}^{(1)}={\cal H}_{A}^{(1)}\otimes {\cal H}_{B}^{(1)},\,\,\,
    {\cal H}^{(2)}={\cal H}_{A}^{(2)}\otimes {\cal H}_{B}^{(2)}.
\end{eqnarray}
For all $\sigma^{(1)}\in {\cal S}({\cal H}^{(1)})$ and
$\sigma^{(2)} \in {\cal S}({\cal H}^{(2)})$ then
\begin{equation}
    E(\sigma^{(1)}\otimes \sigma^{(2)})=E(\sigma^{(1)})+
    E(\sigma^{(2)}).
\end{equation}
\end{itemize}
For a thorough discussion of these properties, see Refs. \cite{Horo1,Uniqueness}.
Functionals with the properties (i)-(iv) will as usual be denoted
as entanglement montones.

\section{Properties of $E_A$}

The first statement that we will prove is the
property of $E_A$ to be an entanglement monotone in the abovementioned sense,
the second will be the strong additivity property.\\

\noindent
{\bf Proposition 1.}
{\it  $E_A:{\cal S}({\cal H})\longrightarrow
\reals^{+}\cup \{\infty\}$
with
\begin{equation}
    E_A(\sigma):=
    \inf_{\rho \in {\cal D}_{\sigma}({\cal H})}
    S(\rho||\sigma).
\end{equation}
 has the  properties (i)-(iv), i.e., it is
an entanglement monotone.}\\

{\it Proof.} Properties (i) and (ii) are obvious from the
definition, given that the relative entropy is not negative for
all pairs of states. Let $\sigma_1,\sigma_2\in {\cal S}({\cal
H})$, and let $\rho_1\in {\cal D}_{\sigma_1}({\cal H})$ and
$\rho_2\in {\cal D}_{\sigma_2}({\cal H})$ be (not uniquely
defined) states that are 'closest' to $\sigma_1$ and $\sigma_2$,
respectively, in the sense that for $i=1,2$
\begin{eqnarray}
    E_A(\sigma_i)= S(\rho_i\|\sigma_i).
\end{eqnarray}
Such states always exist, due to the lower-semicontinuity of the
relative entropy, and due to the fact that the sets ${\cal
D}_{\sigma_1} ({\cal H})$ and
 ${\cal D}_{\sigma_2}
({\cal H})$
are compact. Then,
for any $\lambda\in[0,1]$,
\begin{equation}
    \lambda\rho_1+(1-\lambda)\rho_2
    \in {\cal D}_{\lambda\sigma_1 + (1-\lambda)\sigma_2}({\cal H}).
\end{equation}
The convexity of $E_A$ hence follows from the
joint convexity of the relative entropy, and one obtains
\begin{eqnarray}
    \!\lambda E_A(\sigma_1)\!\!&+&\!\!(1-\lambda) E_A(\sigma_2) \nonumber\\
    & = &
    \lambda S(\rho_1\|\sigma_1)
    + (1-\lambda)
    S(\rho_2\|\sigma_2)\nonumber\\
    &\geq&
    S(\lambda \rho_1+(1-\lambda)\rho_2\|
    \lambda\sigma_1 + (1-\lambda)\sigma_2).
\end{eqnarray}
This is property (iii).
The monotonicity of $E_A$ under
local operations can be shown as follows:
As mixing can only reduce the degree of entanglement as measured
in terms of $E_A$, it is sufficient to prove that
Eq.\ (\ref{mon}) holds with
\begin{eqnarray}
    \sigma_i &:=& (A_i\otimes {\mathbbm{1}}) \sigma (A_i\otimes {\mathbbm{1}})^\dagger/p_i,\\
    p_i &:=& \text{tr}[ (A_i\otimes {\mathbbm{1}}) \sigma (A_i\otimes {\mathbbm{1}})^\dagger],
\end{eqnarray}
where
$A_i$, $i=1,...,N$, are operators satisfying
$\sum_{i=1}^N A_i^\dagger A_i={\mathbbm{1}}$.
Let $\rho\in {\cal D}_\sigma
({\cal H})$ be the state that satisfies
$E_A(\sigma)= S(\rho\|\sigma)$.
The state that is obtained after the measurement
on $\rho$ is given by
\begin{eqnarray}
    \rho_i:=  (A_i\otimes {\mathbbm{1}})  \rho (A_i\otimes {\mathbbm{1}})^\dagger /
    \text{tr}[(A_i\otimes {\mathbbm{1}})  \rho (A_i\otimes {\mathbbm{1}})^\dagger].
\end{eqnarray}
As a consequence of  $\rho\in {\cal D}({\cal H})$ also
\begin{equation}
    \rho_i\in {\cal D}_{ \sigma_i}
    ({\cal H})
\end{equation}
holds for all $i=1,...,N$.
 The Kraus operators act in the Hilbert space of one
party only and therefore,
\begin{eqnarray}
    p_i&=&
    \text{tr}[(A_i\otimes {\mathbbm{1}}) \sigma (A_i\otimes {\mathbbm{1}})^\dagger] \nonumber \\
    &=&\text{tr}[(A_i\otimes {\mathbbm{1}}) \rho (A_i\otimes {\mathbbm{1}})^\dagger].
\end{eqnarray}
This is where the assumption that $\rho\in {\cal D}_\sigma({\cal H})$
enters the proof. Then
\begin{eqnarray}\label{re}
    &&\sum_{i=1}^N p_i S\left( \rho_i || \sigma_i \right) =
    \nonumber \\
    &&\sum_{i=1}^N  \text{tr}[(A_i\otimes {\mathbbm{1}}) \rho (A_i\otimes {\mathbbm{1}})^\dagger ]
    S\left( \rho_i || \sigma_i \right).
\end{eqnarray}
The right hand side of Eq.\ (\ref{re}) can
now be bounded from above by $S(\rho\|\sigma)$,
by virtue of an inequality of Ref.\ \cite{Wehrl}
(see also \cite{Plenio}), i.e.,
\begin{eqnarray}
    \sum_{i=1}^N  \text{tr}[(A_i\otimes {\mathbbm{1}}) \rho
    (A_i\otimes {\mathbbm{1}})^\dagger ]
    S\left( \rho_i ||
    \sigma_i
    \right)
    \leq
    S(\rho||\sigma).\label{dunno}
\end{eqnarray}
Let $\omega_i\in {\cal D}_{\sigma_i}({\cal H})$ be the
state satisfying $E_A(\sigma_i)= S(\omega_i|| \sigma_i)$, then
\begin{eqnarray}
    E_A(\sigma) & =& S(\rho||\sigma)
    \geq  \sum_{i=1}^N p_i
    S\left( \omega_i ||
    \sigma_i
    \right)\nonumber\\
    & =&\sum_{i=1}^N p_i E_A(\sigma_i).
\end{eqnarray}
This is property (iii), the monotonicity under local
operations.\proofend

\noindent
{\bf Proposition 2.} {\it $E_A$  is strongly
additive.}\\

{\it Proof.} Let ${\cal H}$ be a finite-dimensional
Hilbert space with the above product structure
${\cal H}={\cal H}^{(1)}\otimes {\cal H}^{(2)}$,
and let $\rho \in {\cal S}({\cal H})$.
From the conditional expectation property
of the relative entropy \cite{Ohya}
with respect to the partial trace projection
it follows that
\begin{eqnarray*}
  S(\rho||  \sigma^{(1)}  \otimes \sigma^{(2)})=
  S(\text{tr}_2[\rho] || \sigma^{(1)} )
  + S(\rho||\text{tr}_2[\rho]\otimes \sigma^{(2)})
\end{eqnarray*}
for all
$\sigma^{(1)}\in {\cal S}({\cal H}^{(1)})$,
$\sigma^{(2)}\in {\cal S}({\cal H}^{(2)})$,
such  that
\begin{eqnarray}
        S(\rho||\sigma^{(1)}\otimes \sigma^{(2)})&=&
        S(\text{tr}_2[\rho] || \sigma^{(1)})+
    S(\text{tr}_2[\rho] ||\sigma^{(2)})\nonumber\\
    &+&
        S(\rho||\text{tr}_2[\rho]\otimes \text{tr}_1[\rho]),
\end{eqnarray}
and hence
\begin{eqnarray}
    S(\rho||\sigma^{(1)}\otimes \sigma^{(2)})\geq
        S(\text{tr}_2[\rho]\otimes\text{tr}_1[\rho]||
    \sigma^{(1)}\otimes \sigma^{(2)}).
\end{eqnarray}
Moreover, if $\rho\in {\cal D}_{\sigma^{(1)}\otimes \sigma^{(2)}}({\cal H})$
for given $\sigma^{(1)}\in {\cal S}({\cal H}^{(1)})$ and
$\sigma^{(2)}\in {\cal S}({\cal H}^{(2)})$,
then also
\begin{equation}
\text{tr}_2[\rho]\otimes\text{tr}_1[\rho] \in
{\cal D}_{\sigma^{(1)}\otimes \sigma^{(2)}}({\cal H}).
\end{equation}
This in turn implies that any 'closest'
state $\rho\in {\cal D}_{\sigma^{(1)}\otimes \sigma^{(2)}}({\cal H})$
that satisfies $E_A(\sigma^{(1)}\otimes \sigma^{(2)})= S(\rho\| \sigma^{(1)}\otimes \sigma^{(2)})$
can be replaced by
$\text{tr}_2[\rho]\otimes\text{tr}_1[\rho]$,
which again satisfies
\begin{eqnarray}
    \!\!\!\! E_A(\sigma^{(1)}\otimes \sigma^{(2)})\!&=&\!
    S(\text{tr}_2[\rho]\otimes\text{tr}_1[\rho] \| \sigma^{(1)}\otimes \sigma^{(2)})\nonumber\\
    &=& S(\text{tr}_2[\rho] \| \sigma^{(1)})+S(\text{tr}_1[\rho]\|  \sigma^{(2)}).
\end{eqnarray}
Therefore,
\begin{equation}
    E_A(\sigma^{(1)}\otimes \sigma^{(2)})=
    E_A(\sigma^{(1)})+
    E_A(\sigma^{(2)}),
\end{equation}
meaning that $E_A$ is strongly additive.
\proofend

%
%
%

According to the statistical interpretation given in Section VI,
$E_A$ has the property to be divergent for sequences of mixed
states converging to pure states, and hence does not distinguish
pure states in their degree of entanglement.
Therefore, it is not a very
practical measure of entanglement. However,
as it is the only strongly additive entanglement monotone
known to date, it appears fruitful to investigate
the conditional expectation property of the relative entropy
of entanglement further in order to try to
construct strongly additive entanglement monotones that have
the ability to discriminate between the degrees of entanglement
of pure states.
\section{Properties of $E_M$}

In this section we will investigate
the properties of the quantity
$E_M$. First we will show that the relative
entropy of entanglement $E_M$ retains
all properties of an entanglement monotone if
one additionally requires that the closest disentangled state
has the same reductions as the primary state.
This observation implies a simplification when it
comes to actually evaluating the relative entropy of
entanglement, be it with analytical or with numerical means, because
the dimension of the feasible set is smaller.\\

\noindent{\bf Proposition 3.}
{\it $E_M:{\cal S}({\cal H})\longrightarrow
\reals^+$ with
\begin{equation}
    E_M(\sigma)= \inf_{\rho\in {\cal D}_\sigma
    ({\cal H})}
    S(\sigma\|\rho)
\end{equation}
is an entanglement monotone with properties (i)-(iv).}\\

{\it Proof.}
Properties (i), (i), and (iii) can be shown just as before.
Again for states $\sigma, \sigma_1, \sigma_2\in {\cal S}({\cal H})$
and
$
    \rho\in {\cal D}_\sigma({\cal H})$
$\rho_1\in {\cal D}_{\sigma_1}({\cal H})$,
    $\rho_2\in {\cal D}_{\sigma_2}({\cal H})$
it follows that
\begin{equation}
    A\rho A^\dagger/\text{tr}[A \rho A^\dagger]
    \in {\cal D}_{
    A\sigma A^\dagger/\text{tr}[A \sigma A^\dagger]}({\cal H})
\end{equation}
for all $A$, and
\begin{equation}
    \lambda\rho_1+(1-\lambda)\rho_2 \in
    {\cal D}_{\lambda\sigma_1+(1-\lambda)\sigma_2}({\cal H}).
\end{equation}
With the notation of the proof of
property (iv),
\begin{eqnarray}
    E_M(\sigma) & =&  S(\sigma ||\rho)
    \geq
    \sum_{i=1}^N p_i
    S\left( \sigma_i ||
    \omega_i
    \right)\nonumber\\ &=&
    \sum_{i=1}^N p_i E_M(\sigma_i).
\end{eqnarray}
\proofend

Hence, the relative entropy of entanglement is still
an entanglement monotone when one restricts the
set of feasible PPT or separable states to those
that are locally identical to a given state.
At first it does not even seem obvious that
$E_M$ is even different from the original relative
entropy of entanglement. In fact, all states $\sigma$
considered in Ref.\ \cite{Plenio} satisfy
\begin{equation}
E_M(\sigma) = \inf_{\rho\in {\cal D}({\cal H})}S(\sigma\|\rho).
\end{equation}
Also, for all $UU$ and $OO$-symmetric 
states the two quantities are obviously the same. This
version of the relative entropy of entanglement is strictly
sub-additive, just as the relative entropy of entanglement with
respect to the unrestricted sets of separable states or PPT
states. However -- on the basis of numerical studies --
it turns out that the two quantities are not
identical general, 
and that there exist states for which the two
entanglement measures do not give the same value \cite{S1}.
This means that
the disentangled state that can be least distinguished from a
given primary state may have the property that it can already be
locally distinguished. \\

\noindent {\bf Example 4.} We have numerically evaluated
the difference $E_R(\rho_p)-E_M(\rho_p)$ 
between the (ordinary) relative entropy of
entanglement $E_R$ and the modified quantity $E_M$ for
states on ${\mathbbm{C}}^2 \otimes {\mathbbm{C}}^2$
of the form
\begin{equation}
	\rho_p:= p |\psi\rangle\langle\psi|+(1-p) {\mathbbm{1}}/4,\,\,\,\,
	p\in[0,1],
\end{equation}
where 
\begin{equation}
	|\psi\rangle:= 
	\left(
	|0,0\rangle + (1+i) |0,1\rangle+ (1-i) 
	|1,0\rangle
	\right)/5^{1/2}.
\end{equation}
Figure 1 shows this difference  $E_R(\rho_p)-E_M(\rho_p)$ as
a function of $p\in[0,1]$. The difference is in fact quite small, but
significant, given the accuracy of the program \cite{Num}. 
Numerical studies indicate that differences of this order
of magnitude are typical for generic quantum states on
${\mathbbm{C}}^2 \otimes {\mathbbm{C}}^2$.

\begin{figure}
\centerline{
        \epsfxsize=8.5cm
       \epsfbox{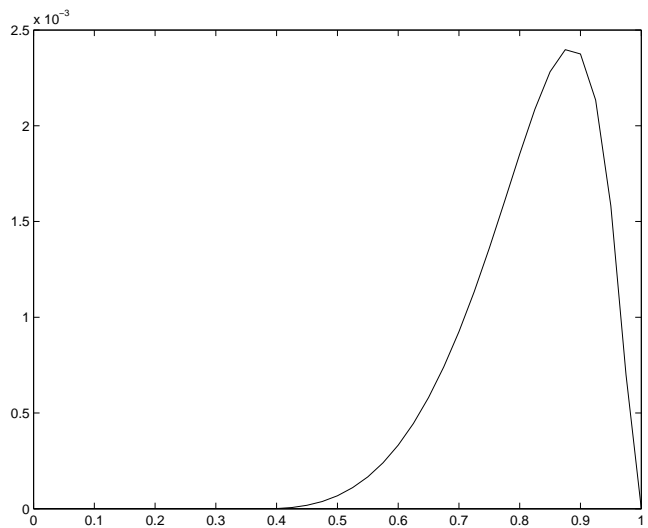}
}

\vspace{.2cm}
\caption{The difference $E_R(\rho_p)-E_M(\rho_p)$ for the state
$\rho_p$
as a function of $p$.}
\end{figure}

\section{Properties of $E_T$}

We now turn to the third quantity $E_T$, the minimal distance of a
state $\sigma$ to the set ${\cal D}_\sigma({\cal H})$ with respect
to the trace-norm difference. We will show that also this quantity
is a proper measure of entanglement. Other physically interesting
quantities of this type have been considered in the literature, in
particular, the minimal Hilbert-Schmidt distance of a state to the
set of PPT states  \cite{HS1,HS,Witte}. For the latter quantity
the resulting minimization problem can in fact be solved
\cite{HS1}. However, then the resulting quantity is unfortunately
no proper entanglement measure \cite{NotHS}.\\


\noindent{\bf Proposition 5.} {\it  $E_T:
{\cal S}({\cal H})\longrightarrow
\reals^+$ with
\begin{equation}
    E_T(\sigma)=\min_{\rho\in{\cal D}_\sigma({\cal H})}\|\sigma-\rho\|_1
\end{equation}
is an entanglement monotone with properties (i) - (iv).}\\

{\it Proof.}
Clearly, $E_T(\rho)=0$ for a state $\rho\in{\cal D}({\cal H})$.
In order to show convexity one can proceed
just as in the proofs of Proposition 1 and 3:
the convexity then follows from the triangle inequality
for the trace norm.
The remaining task is to show that
it is monotone under local operations.
Again,
\begin{eqnarray}
p_i &=& \text{tr}[(A_i\otimes {\mathbbm{1}}) \rho
(A_i\otimes{\mathbbm{1}})^\dagger] \\ &=&
\text{tr}[(A_i\otimes {\mathbbm{1}}) \sigma (A_i\otimes
{\mathbbm{1}})^\dagger] \nonumber
\end{eqnarray}
for all
$\rho\in {\cal D}_{\sigma}({\cal H})$, and
$   (A_i\otimes {\mathbbm{1}})\rho(A_i\otimes {\mathbbm{1}})^\dagger/p_i
    \in {\cal D}_{\sigma_i}$.
Hence,
\begin{eqnarray}
&&\!\!\!\sum_{i=1}^N  p_i E_T( \sigma_i) = \\
&&\!\!\!\sum_{i=1}^N p_i \min_{\rho_i \in{\cal D}_{\sigma_i}({\cal
H}) }
    \|(A_i\otimes {\mathbbm{1}}) \sigma (A_i\otimes {\mathbbm{1}})^\dagger/p_i -
    \rho_i\|_1,\nonumber
\end{eqnarray}
and since
\begin{eqnarray}
    &\min\limits_{\rho_i \in{\cal D}_{\sigma_i}({\cal H}) }&
    \|(A_i\otimes {\mathbbm{1}}) \sigma (A_i\otimes {\mathbbm{1}})^\dagger/p_i -
    \rho_i\|_1\\
    \leq
    &\min\limits_{\rho \in{\cal D}_\sigma({\cal H}) }&
    \frac{\|(A_i\otimes {\mathbbm{1}})\sigma (A_i\otimes {\mathbbm{1}})^\dagger -
    (A_i\otimes {\mathbbm{1}})\rho (A_i\otimes {\mathbbm{1}})^\dagger\|_1}{p_i},\nonumber
\end{eqnarray}
we arrive at
\begin{equation}
    \sum_{i=1}^N  p_i E_T( \sigma_i)\leq\min_{\rho \in{\cal D}_\sigma({\cal H}) } \sum_{i=1}^N
        \|(A_i\otimes {\mathbbm{1}}) (\sigma-\rho) (A_i\otimes {\mathbbm{1}})^\dagger\|_1.
\end{equation}
Property (iv) then follows from Lemma 6 (presented below), which
yields
\begin{eqnarray}
    \sum_{i=1}^N  p_i E_T( \sigma_i)
        &\leq&\min_{\rho \in{\cal D}_\sigma({\cal H}) }  \sum_{i=1}^N
        \|(A_i\otimes {\mathbbm{1}})^\dagger (A_i\otimes {\mathbbm{1}})
    |\sigma-\rho|\, \|_1\nonumber\\
    &\leq&
    \min_{\rho \in{\cal D}_\sigma({\cal H}) }  \sum_{i=1}^N
    \text{tr}[(A_i\otimes {\mathbbm{1}})^\dagger (A_i\otimes {\mathbbm{1}})
    |\sigma-\rho|\
    ]\nonumber\\
    &=&
     \min_{\rho \in{\cal D}_\sigma({\cal H}) }
    \|\sigma-\rho\|_1=E_T(\sigma).
\end{eqnarray}
Hence, $E_T$ is monotone under local operations.
\proofend

\noindent{\bf Lemma 6.} {\it Let $A,B$ be complex
$n\times n$ matrices,
and assume that $B=B^\dagger$. Then
\begin{equation}
    \| A B A^\dagger \|_1 \leq \| A^\dagger A |B| \|_1\label{Helpful}
\end{equation}
holds.}\\

{\it Proof.}
The trace norm$\|.\|_1$
is a unitarily invariant norm,
and $A B A^\dagger$ is a normal matrix \cite{Bhatia}.
Hence
\begin{equation}
    \| A (B A^\dagger) \|_1\leq  \|  (B A^\dagger) A\|_1
\end{equation}
(see Ref.\ \cite{Bhatia}), and therefore,
\begin{eqnarray}
\|(B A^\dagger) A\|_1 &=&
\text{tr}[(A^\dagger A  B^\dagger B A^\dagger A )^{1/2}] \nonumber \\
&=& \text{tr}[(A^\dagger A |B|^2 A^\dagger A )^{1/2}]\nonumber \\
&=& \| A^\dagger A |B| \|_1,
\end{eqnarray}
which gives rise to Eq.\ (\ref{Helpful}).
\proofend

\smallskip
Hence, $E_T$ is a proper entanglement monotone, yet it does not
exhibit an additivity property, and it is not asymptotically
continuous on pure states. It should be noted that the weaker
condition $E_T({\cal E}(\sigma)) \leq  E_T( \sigma)$ for all
trace-preserving maps ${\cal E}$ corresponding to local operations
with classical communication and all states $\sigma$ follows
immediately from the fact that the trace norm fulfills
\begin{equation}
    \| {\cal E}(\sigma)- {\cal E}(\rho)\|_1
    \leq \|\sigma-\rho\|_1
\end{equation}
for all trace-preserving
completely positive maps  ${\cal E}$ and all
states $\sigma,\rho$. The Hilbert-Schmidt norm in turn
does not have this property \cite{NotHS}.

\section{Distance measures and state distinguishability}

In this section we will give an interpretation of the
three quantities $E_A$, $E_M$ and $E_T$
in terms of hypothesis testing.
The problem of distinguishing quantum mechanical states
can be formulated as testing two competing claims, see Refs.\ \cite{Hiai,Stein,Fuchs}.
In this setup one considers a single dichotomic generalized measurement acting on a state that is known to
be either $\omega$ or $\xi$, with equal a priori probabilities. The measurement is
represented by two positive operators $E$ and
${\mathbbm{1}}-E$, with $E$ satisfying $0\leq E\leq {\mathbbm{1}}$.
On the basis of the outcome of the measurement
one can then make the decision to accept either the hypothesis that the state $\omega$
has been prepared (the null-hypothesis), or the hypothesis that
the state $\xi$ has been prepared (the alternative hypothesis).
The error probabilities of first and second kind related to this decision are
given by
\begin{eqnarray}
    \alpha(\omega,\xi;E) &:=& \text{tr}[ \omega({\mathbbm{1}}-E)],\\
    \beta(\omega,\xi;E)  &:=& \text{tr}[\xi E].
\end{eqnarray}

The trace-norm difference of the two states $\omega$ and $\xi$
can be written in terms of these error probabilities as follows.
According to the variational characterisation of the trace norm,
\begin{equation}
    \|\omega -\xi \|_1 = \max_{X,\|X \|\leq 1} \text{tr}[(\omega-\xi) X],
\end{equation}
where $\|.\|$ denotes the standard operator norm \cite{Bhatia}.
There is a one-to-one relation between the allowed $X$ appearing here and the set of hypothesis
tests:
$E=(X+{\mathbbm{1}})/2$. Hence, $\text{tr}[(\omega-\xi)X] = 2 \text{tr}[(\omega-\xi)E] $
implying that the quantity $E_T$ can be interpreted as
\begin{equation}
    E_T(\sigma) = 2 \inf_{\rho\in{\cal D}_\sigma({\cal H})}
    \max_E \left( 1-\alpha(\sigma,\rho;E) - \beta(\sigma,\rho;E) \right),
\end{equation}
with $E$ any test ($0\le E\le {\mathbbm{1}}$).
Due to the restriction $\rho \in{\cal D}_{\sigma}({\cal H})$,
one compares the primary state $\sigma$ only with those separable
(PPT) $\rho$ that have the same reductions as $\sigma$. Clearly,
tests consisting of tensor products $E=E_A\otimes E_B$ cannot
distinguish such states at all, as the outcomes will exhibit the
same probability distributions for both states. 

The quantum hypothesis tests related to $E_T$ are restricted to a
single measurement on a single bi-partite quantum system. The
quantities $E_M$ and $E_A$ can in some sense be considered the
asymptotic analogues of $E_T$. The connection between the relative
entropy and the error probabilities in quantum hypothesis testing
has been thoroughly discussed in Refs.\ \cite{Hiai,Stein,Fuchs}.
In the asymptotic setting one considers sequences consisting of
tuples of $n$ identically prepared states, $\omega^{\otimes n}$
and $\xi^{\otimes n}$, 
and a sequence of tests $\{E_n\}_{n=0}^\infty$, 
where
$0\leq E_n\leq {\mathbbm{1}}$ and $E_n$ operates on an $n$-tuple. To
every test in the sequence, one can again ascribe two error
probabilities:
\begin{eqnarray}
    \alpha_n(\omega,\xi;E_n) &:=&\text{tr}[\omega^{\otimes n} ({\mathbbm{1}}-E_n)] \\
    \beta_n(\omega,\xi;E_n) &:=& \text{tr}[\xi^{\otimes n} E_n].
\end{eqnarray}
For any $\varepsilon>0$ define \cite{Stein}
\begin{eqnarray}
&&    \beta^*_n(\omega, \xi; \varepsilon) := \nonumber\\
&&    \min \{\beta_n (E_n):
    0\leq E_n\leq {\mathbbm{1}},
    \alpha_n(\omega, \xi;E_n)<\varepsilon\}.
\end{eqnarray}
It has been shown \cite{Stein}
that for any $0\leq \varepsilon <1$
\begin{equation}\label{h}
    \lim_{n\rightarrow \infty} \frac{1}{n} \log
    \beta^*_n(\varepsilon) = -S(\omega\|\xi).
\end{equation}
This means that if one requires that the error probability of
first kind is no larger than $\varepsilon$, then the error
probability of second kind goes to zero according to Eq.\
(\ref{h}). Having this in mind, the quantity $E_M$ can be
interpreted as an asymptotic measure of distinguishing $\sigma\in
{\cal D}({\cal H})$ from the closest $\rho\in {\cal
D}_\sigma({\cal H})$ with the same reductions as $\sigma$. In
turn, $E_A$ is a similar quantity but with the roles of $\sigma$
and $\rho$ reversed. The asymmetry comes from the asymmetry of the
roles of the error probabilities of first and second kind.

Note that, within this interpretation, the divergence of $E_A$
on pure states becomes plausible. If $\xi$ is pure,
choosing the sequence of tests  $\{E_n\}_{n=0}^\infty$ with

\begin{equation}
	E_n := {\mathbbm{1}}-\xi^{\otimes n}
\end{equation}
yields a $\beta_n$ equal to zero for any $n$ (this can only happen for pure $\xi$)
and an $\alpha_n$ equal to $\text{tr}[\omega\xi]^n$, which always becomes
smaller than any chosen value of $\varepsilon>0$ from some {\it finite} value of $n$
onwards (that is, presuming $\omega\neq\xi$).
Hence, for any choice of $\varepsilon$ there is a finite value
of $n$, say $n(\varepsilon)$, such that $\beta^*_n(\varepsilon)=0$
for $n\ge n(\varepsilon)$. Asymptotical convergence of $\beta^*_n(\varepsilon)$
is therefore faster than exponential so that
$\{ \log \beta^*_n(\varepsilon)/n\}_{n=1}^\infty$ tends to
minus infinity.

\section{Summary and Conclusion}

In this paper we have investigated three variants of the relative
entropy of entanglement, all three of which can be related to the
problem of distinguishing a primary state from the closest
disentangled or PPT state that has the same reductions as the
primary state. This approach was motivated by the desire to flesh
out the genuinely non-local distinguishability of a primary state
from the closest disentangled state. The three functionals have
been found to be legitimate measures of entanglement.
Additionally, one functional has the property of being strongly
additive, thereby showing that monotonicity, convexity and strong
additivity are compatible in principle. This additivity
essentially originates from the conditional expectation property
of the relative entropy. In the light of this observation it
appears interesting to further study the implications of the
conditional expectation property of the relative entropy on
quantum information theory.

\begin{acknowledgments}
We would like to thank M.\ Horodecki,
R.F.\ Werner, and M.\ Hayashi for helpful remarks.
This work has been supported by the EQUIP project of the European
Union, the Alexander-von-Humboldt Foundation, the EPSRC and the
ESF programme for ''Quantum Information
Theory and Quantum Computation''. 
\end{acknowledgments}



\begin{thebibliography}{99}
\bibitem{PPT}
    Under PPT operations, that is, quantum operations preserving the
    positivity of the partial transpose, it is an open question whether
    there exists a unique measure of entanglement \cite{Asy3,Jon}. 
    In fact, there
    exist truly mixed states for which asymptotic state manipulation
    under PPT operations can be shown to be
    reversible \cite{Asy3}, which points towards the possibility of
    having a unique measure of entanglement under PPT operations.
\bibitem{S1}
Ref.\ \cite{S} presents a set of equations that has
to be satisfied for the closest state to have identical
reductions as the primary state in the relative entropy of
entanglement.
\bibitem{Num}
	The algorithm that has been used to numerically evaluate the two 
	quantities will be
	discussed elsewhere.
\end{thebibliography}
\end{document}